\documentstyle[11pt, epsf, a4]{article}

\newcommand{\dquad}{\qquad\qquad}

\newcommand{\gl}{\;\;=\;\;}
\newcommand{\gr}{\;\;>\;\;}
\newcommand{\kl}{\;\;<\;\;}
\newcommand{\Sim}{\;\;\sim\;\;}

\newcommand{\pl}{\;+\;}
\newcommand{\mi}{\;-\;}

\newcommand{\be}{\begin{equation}}
\newcommand{\ee}{\end{equation}}
\newcommand{\bea}{\begin{eqnarray}}
\newcommand{\eea}{\end{eqnarray}}

\newcommand{\bc}{\begin{center}}
\newcommand{\ec}{\end{center}}

\newcommand{\bt}[1]{\begin{tabular}{#1}}
\newcommand{\et}{\end{tabular}}

\renewcommand{\(}{\left(}
\renewcommand{\)}{\right)}
\renewcommand{\[}{\left[}

\newcommand{\eqn}[1]{(eq.~(\ref{#1}))}
\newcommand{\eq}[1]{eq.~(\ref{#1})}

\newcommand{\ph}{\varphi}
\newcommand{\del}{\partial}
\newcommand{\laT}{\lambda_T}
\newcommand{\Tr}{{\rm Tr}\,}

\newcommand{\calO}{{\cal O}}
\newcommand{\ra}{\rightarrow}

\def\mW{m_W}
\def\mWq{m_W^2}
\def\mWf{m_W^4}
\def\mGs{m_{\rm Gs}}
\def\mGsq{m_{\rm Gs}^2}
\def\mGsf{m_{\rm Gs}^4}
\def\mH{m_{\rm H}}

\def\lsim{\mbox{\raisebox{0.05cm}[0cm][0cm]{$<$}\hspace{-0.42cm}
          \raisebox{-0.12cm}[0cm][0cm]{$\sim$}}}
\def\gsim{\mbox{\raisebox{0.05cm}[0cm][0cm]{$>$}\hspace{-0.42cm}
          \raisebox{-0.12cm}[0cm][0cm]{$\sim$}}}

\let\3=\ss

\begin{document}

\begin{titlepage}
\begin{flushright}
HD--THEP--95--53\\
hep-ph/9512340\\
December 18, 1995\\
\end{flushright}
\vspace{1.5cm}
\begin{center}
\mbox{\bf\LARGE Perturbative Contributions to the}\\
\vspace{.2cm}
{\bf\LARGE  Electroweak Interface Tension}\\
\vspace{1cm}
{\bf
Jochen Kripfganz\footnote{supported by Deutsche
Forschungsgemeinschaft}\\
\vspace{.3cm}
Andreas Laser\footnote{supported by Landesgraduiertenf\"orderung
Baden-W\"urttemberg\\[1ex]
e-mail addresses:\\[0.5ex]
\begin{tabular}{r}
J.Kripfganz@thphys.uni-heidelberg.de\\
    A.Laser@thphys.uni-heidelberg.de\\
M.G.Schmidt@thphys.uni-heidelberg.de
\end{tabular}}\\
\vspace{.3cm}
Michael G.~Schmidt\\
}
\vspace{1cm}
Institut  f\"ur Theoretische Physik\\
Universit\"at Heidelberg\\
Philosophenweg 16\\
D-69120 Heidelberg, FRG\\
\vspace{1.5cm}
{\bf Abstract}\\
\end{center}
The main perturbative
contribution to the free energy of an electroweak
interface is due to the effective potential and the tree
level kinetic term. The derivative corrections are investigated
with one-loop perturbation theory. The action is
treated in derivative, in heat kernel, and in a multi local
expansion. The massive contributions turn out to be well described
by the $Z$-factor. The massless mode, plagued by infrared problems,
is numerically less important. Its perturbatively reliable part
can by calculated in derivative expansion as well.
A self consistent way to include the $Z$-factor in the formula for
the interface tension is presented.
\end{titlepage}

\section{Introduction}

The electroweak phase transition is under intense
investigation \cite{ref1}-\cite{Wir3}
mostly because of possible effects in the early universe, in particular
baryon number generation. Besides of this, it is the prototype of a
phase transition in a gauge theory where careful understanding will
be useful also in other settings, e.g.\ if the standard model electroweak
phase transition should turn out to be too weakly first order for
spectacular effects, and modified theories have to be considered.

The electroweak effective potential at high temperatures has been treated
carefully in two loop order of perturbation theory
\cite{ArnoldEs}-\cite{Wir2}.
Comparison with lattice nonperturbative results
\cite{KajantieEA2, FodorEA2,CsikorEA,IlgenfEA} gives good agreement
for the Higgs and $W$-boson masses in the broken phase up to
zero temperature Higgs masses of about 70 GeV.
The hot electroweak phase appears as a case of a
confining gauge system.
The Higgs and $W$-correlation masses have recently been
explained by a bound state model within some
approximations \cite{Wir3}.
The critical temperature and
the latent heat are expected to
be influenced by nonperturbative effects
in the hot (symmetric) phase. Two-loop perturbation theory agrees,
nevertheless, surprisingly well with the values of these quantities
measured on the lattice.

The situation for the interface tension is quite different.
It is more difficult to measure
on the lattice.
While one lattice group finds good agreement \cite{CsikorEA}
another group reports large deviations at intermediate
Higgs masses \cite{KajantieEA2}.
Furthermore the perturbative values of the interface tension calculated
from
\be
\tilde{\sigma} \gl \int^{\varphi_A}_{\varphi_S}\!\! d\varphi\,
            \sqrt{2\, V(\varphi(r))}
\ee
in the literature do not include important perturbative effects,
as we will argue in this paper.

The interface tension is the free energy per area of interface between
the hot and the cold electroweak phase. It is identical to the
surface tension of thin walled electroweak critical bubbles
which trigger the onset of the phase transition close to the
critical temperature. The latter is read off from the part of the
bubble free energy proportional to  $4\pi R^2$ ($R$=radius).

In this paper we discuss the modification of the interface tension
by one loop perturbation theory for the derivative terms
of the action. In particular we address the question
of whether these modification can be understood just by a $Z(\ph)$-factor
in front of the kinetic term $\partial\ph\partial\ph$ or whether
all the other derivative terms
together give a sizable effect on the interface tension.

The paper is organized as follows.
In chapter \ref{HTPT} we recapitulate the one-loop high temperature
perturbation theory. We especially address the one-loop $Z$-factor.
In chapter \ref{HODT} the effective action is expanded in a
multi local way (section \ref{MLE}) and with the heat kernel expansion
(section \ref{HKE}).
The connection between both expansions and the derivative expansion
is clarified in section \ref{CDMLHKE}.
The infrared problems of the massless mode are investigated in
chapter \ref{MM}.  We show that the divergent
heat kernel operators can be resummed to give a finite contribution
(section \ref{RIDHKO}).
The derivative expansion of these modes is
treated in detail (section \ref{IPDE}).
In chapter \ref{IT} a self consistent way to include the $Z$-factor
in the formula for
the interface tension is presented.
Chapter \ref{Conclusion} gives our conclusions.

\section{High Temperature Perturbation Theory}\label{HTPT}

The electroweak standard theory  at the transition temperature
allows the high tem\-pera\-ture
expansion. To
good accuracy it can be described by a local
effective 3-dimensional
$SU(2)$-Higgs-model whose parameters can be calculated from the fundamental
4-dimen\-sion\-al theory \cite{KajantieEA1}.
All infrared problems appearing in the unbroken phase are
covered by this effective theory.
The nature of the phase transition depends substantially
on the static Matsubara frequencies.
Calculating the  one-loop effective action of a scalar background field
in the derivative expansion using a general
't~Hooft background gauge with gauge-fixing
parameter $\xi$ one gets \cite{Wir1,Wir2}
\be\label{Seff}
S_{\rm eff}[\varphi] \gl  \frac{1}{g_3(T)^2}
        \int d^3r \left[V_{\rm eff}(\varphi)
     + \frac{1}{2} Z_{\rm H}(\varphi) \partial_i\varphi \partial_i\varphi
     + {\cal O}(\partial^4)\right]
\ee
with the effective potential
\bea
V_{\rm eff}(\varphi) &=& V_{\rm ht}(\varphi) \;-\; \frac{g_3(T)^2}{12 \pi}
          \left( 9 m_W^3 - 6 m_{\rm gh}^3 +
                 3 m_{\rm Gs}^3 + 3 m_{\rm H}^3 \right) \label{Veff}\\
V_{\rm ht}(\ph) &=&
   \frac{\lambda_T}{g^2}\left( \frac{\ph^4}{4} -
   \left(\frac{v_0(T)}{v}\right)^2\frac{\ph^2}{2} \right) \label{Vht}
\eea
and the $Z$-factor
\begin{eqnarray}\label{Zfkt}
Z_{\rm H}(\varphi) &=& 1 \;+\;  \frac{g_3^2}{4 \pi} \; \Bigg\{
  -\; \frac{3}{m_{\rm Gs}+m_W} \;
   +\;  \frac{1}{m_W^2} \left(
                \frac{m_{\rm Gs}^3-m_W^3}{m_{\rm Gs}^2-m_W^2} \;- \;
                \frac{m_{\rm Gs}^3-\xi^{3/2}m_W^3}{m_{\rm Gs}^2-\xi m_W^2}
        \right) \nonumber\\
 &&\qquad\qquad  +\;\frac{10 - 13\sqrt{\xi} + 9\xi}{16\,(1 + \sqrt{\xi})}
\; \frac{1}{m_W^3}
   \left(\frac{\partial m_W^2}{\partial \varphi} \right)^2
 \;-\; \frac{2}{16}\, \frac{1}{m_{\rm gh}^3}
      \left(\frac{\partial m_{\rm gh}^2}{\partial \varphi} \right)^2
 \nonumber\\
&&   \qquad\qquad +\; \frac{1}{16}\, \frac{1}{m_{\rm Gs}^3}
      \left(\frac{\partial m_{\rm Gs}^2}{\partial \varphi} \right)^2
    \;+\; \frac{1}{48}\, \frac{1}{m_{\rm H}^3}
      \left(\frac{\partial m_{\rm H}^2}{\partial \varphi} \right)^2
\quad\Bigg\} \quad.
\end{eqnarray}
The squared high temperature Higgs, Goldstone, $W$-boson and ghost masses are
\begin{eqnarray}
m_{\rm H}^2 &=& \frac{\lambda_T}{g^2}
      \(3\varphi^2 - \left(\frac{v_0(T)}{v}\right)^2\)
\quad,  \qquad \qquad\qquad\quad
m_W^2 \gl \frac{1}{4} \varphi^2 \quad, \nonumber \\
m_{\rm Gs}^2 &=&
\frac{\lambda_T}{g^2} \(\varphi^2 - \left(\frac{v_0(T)}{v}\right)^2\)
       \,+\, \frac{1}{4} \xi \varphi^2 \quad,
\qquad\qquad m_{\rm gh}^2 \gl \frac{1}{4} \xi \varphi^2 \quad. \label{masses}
\end{eqnarray}
$\laT$ and $v_0(T)$ can be calculated from the fundamental 4-dimensional
parameters. We identify
\be\label{mH}
\frac{\laT}{g^2} \gl \frac{\bar{m}_{\rm H}^2}{8\,m_W(T\!=\!0)^2}\quad.
\ee
The mass $\bar{m}_{\rm H}$ is of order of the zero temperature
Higgs mass.
$v_0(T)$ is the temperature dependent ``high temperature tree level''
vacuum expectation value of the complex scalar field (cf.~\eq{Vht}).
$v$ was used to rescale fields and will be chosen to
be the field value at the broken minimum later on.
$\,g_3(T)^2 = g T/v \,$ is the rescaled dimensionless
coupling constant of the 3-dimensional theory.
In this respect we differ from the notation $~g_3^2=g^2T~$
sometimes being used.

The two-loop effective potential has been calculated by different
methods \cite{ArnoldEs}-\cite{Laine}.
We use the general 't~Hooft background gauge potential
obtained in \cite{Wir2}.

The different contributions to $Z_{\rm H}(\varphi)$ \eqn{Zfkt}
are due to different graphs.
The first two terms in the  brackets
originate from  a mixed $W\!$-Gs-loop. The other
corrections are generated by a pure $W$-boson, ghost, Goldstone, or Higgs loop
respectively.
The use of the derivative expansion is limited for several reasons:

First of all $Z_{\rm H}(\varphi)$ diverges as $\varphi$ goes to 0,
i.e.\ in the symmetric phase, due to the massless $W$-boson. At this order
of derivative and loop expansion the effective action (\ref{Seff})
is nevertheless finite, since a bubble solution vanishes exponentially in
this phase.
For dimensional reasons the next order of the derivative expansion
behaves like $~{m_W}^{-5} (\del\ph)^4~$ for small $~\ph$.
The two-loop order is expected to give a $~{m_W}^{-2} (\del\ph)^2~$
and a $~{m_W}^{-2}\ln(m_W) (\del\ph)^2~$ term.
All these terms lead to a diverging action. Thus
perturbation theory breaks down. Note that even the effective potential
diverges at four-loop order for $\ph\!\ra\! 0\,$.
The divergences are cured by a finite $W$-boson mass in the symmetric phase
due to confining forces.

Second, $Z_{\rm H}(\varphi)$ becomes negative in some $\ph$-range
depending on $\xi$ (e.g.~for \mbox{$\bar{m}_{\rm H}\gsim 45$GeV}
at $\xi=1$).
This is due to the
numerical dominance of the mixed $W\!$-Gs-loop (cf.~section \ref{MLE}).
Here again
perturbation theory in connection with the derivative expansion
is in bad shape, since the one loop
contribution is larger than the tree part.
In addition a negative $Z$-factor gives rather unphysical results
and no critical bubbles at all.

Finally $Z_{\rm H}(\varphi)$ depends strongly on the gauge-fixing parameter
even in the broken phase,
as we have shown in ref.~\cite{Wir2}.
Therefore higher derivative terms and/or
the two-loop order are expected to give important contributions
to $Z_{\rm H}(\varphi)$.
The gauge-fixing dependence of the potential
diminishes from one to two-loop, and it may be hoped that the same happens
with the $Z$-factor.

\section{Higher Order Derivative Terms}\label{HODT}

It is an obvious question how important the
${\cal O}(\partial^4)$-terms in \eq{Seff} are.
Calculating the next order in the derivative expansion would not by
a feasible way to investigate this question because of the
divergence in the symmetric phase due to massless modes.
One has to use different methods to expand the effective action.
In ref.~\cite{Wir1} we treated the Higgs fluctuation separately
using the heat kernel method
but we did not calculate the contribution of the $W$-boson and
the Goldstone fluctuation to the interface tension.
In this paper we will include these contribution.
We use two different methods of expansion: the heat kernel method
and a multi local expansion.
To the required order for both methods the
't~Hooft-Feynman background gauge (i.e.~$\xi\!=\!1$)
is more convenient. We therefore
restrict ourselves in this chapter to this gauge-fixing.

The one-loop effective action including all derivative corrections is
\begin{equation}
S_{\rm eff}[\varphi] \gl S_{\rm ht}[\varphi] \pl \delta S[\varphi]
\end{equation}
with the high temperature tree level part
\bea \label{Shtbg}
S_{\rm ht} &=& \frac{1}{g_3(T)^2}  \int d^3r \left[\;
\frac{1}{2} \partial_i\varphi \partial_i\varphi \pl
 V_{\rm ht}(\varphi^2) \;\right] \label{Sht}
\eea
and the one-loop correction
\begin{equation}\label{delS}
\delta S \gl \frac{1}{2}\,\ln \det(-\partial^2 + U_{13}) \mi
                     \ln \det (-\partial^2 + U_{\rm gh}) \quad.
\end{equation}
where $U_{13}$ is the 13$\times$13-matrix
\be\label{MflU}
U_{13} \gl
\left(
 \begin{array}{cccc}
   U_4 & 0 & 0 & 0 \\
   0 & U_4 & 0 & 0 \\
   0 & 0 & U_4 & 0 \\
   0 & 0 & 0 & m_{\rm H}^2
 \end{array}
\right)
\qquad
U_4 \gl
\left(
  \begin{array}{cccc}
     m_W^2 & 0 & 0 & \partial_1\varphi \\
     0 & m_W^2 & 0 & \partial_2\varphi \\
     0 & 0 & m_W^2 & \partial_3\varphi \\
     \partial_1\varphi & \partial_2\varphi & \partial_3\varphi & m_{\rm Gs}^2
   \end{array}
\right)
\ee
and the 3$\times$3-ghost matrix is $~m_{\rm gh}^2\, 1_{3 \times 3}$.
The diagonal elements are the finite temperature squared
Higgs, Goldstone, $W$-boson
and ghost masses \eqn{masses}.

We proceed now in choosing a typical interface profile which depends
of course on $\laT$ (i.e.~on the zero temperature Higgs and top mass).
Then the determinants in \eq{delS} are evaluated numerically.
The contributions of $\delta S$ to the effective potential
are removed. The remaining one-loop correction is compared with
the contribution of the $Z$-factor.

A typical interface between the symmetric phase ($\ph\!=\!0$)
and the broken phase ($\ph\!=\!1$)
is given by the thin wall configuration:
\be\label{interface}
\ph(z) \gl \frac{1}{2} \left( 1- \tanh\left(\frac{z-z_0}{d}\right)\right)
\ee
The wall thickness $d$ depends on the quartic coupling $\sqrt{\laT/g^2}$.
At the critical temperature its tree+$\ph^3$-level
value in units of $\,(g\,v)^{-1}\,$ is given by
\be\label{d}
d \gl \frac{\sqrt{8}}{\sqrt{\laT/g^2}} \gl \frac{8\sqrt{\pi}}{\sqrt{g_3^2}}
\quad.
\ee
For small $\laT/g^2$ an expansion in $d^{-1}$ seems to be appropriate.
$d$ may be considered as a variational parameter.

The interface \eq{interface} does not depend on the $x$- or
$y$-coordinate of $\vec{r}$. The matrix $U_{13}$ of \eq{MflU} simplifies
and is partly canceled by the ghost matrix.
The one-loop
effective action \eqn{delS}  becomes
\begin{equation}\label{delS2}
\delta S \gl \frac{3}{2}\,\ln \det(-\partial^2 + U_2) \pl
             \frac{1}{2}\,\ln \det (-\partial^2 + m_{\rm H}^2) \quad.
\end{equation}
with the 2$\times$2-matrix
\be\label{M2}
U_2 \gl
\left(
  \begin{array}{cc}
     m_W^2 & \partial_z\varphi \\
     \partial_z\varphi & m_{\rm Gs}^2
   \end{array}
\right)
\ee

In order to evaluate  the one-loop action \eqn{delS2} approximately
one may choose several different directions to go.\\
A) The derivative expansion discussed above. There is the obvious
problem of infrared divergences blowing up the integrated action
if one expands beyond the $Z$-factor.\\
B) The Barvinsky-Vilkovisky expansion \cite{BaVi} where one fixes the number
of fields and sums over all derivatives, just opposite to the derivative
expansion, obtaining form factors.\\
C) The heat kernel expansion where derivatives and fields are expanded
according to their mass dimension and hence mixed together.
This approach is in between (A) and (B).

An accessible way to treat (B) will be given in the next section \ref{MLE}.
This multi local expansion will automatically sum all derivative
terms for a given power of fields.
After this, we will study the heat kernel
expansion in section \ref{HKE}.

\subsection{The Multi Local Expansion} \label{MLE}

The one-loop contributions to the effective action have the general
structure  (cf.\ eq.~(\ref{delS2})):
\be\label{Gamma}
\Gamma \gl \frac{1}{2}\, \ln \det K  \gl \frac{1}{2}\, \Tr\ln K
\ee
where the matrix $K$ is
$~\(-\del^2+U(\vec{r}\,)\)~$. $U(\vec{r}\,)$ is either $~U_2$
or the  $1\times 1$-matrix
$\left( m_{\rm H}^2 \right)$ in our case but may be different in
other settings.
We split $K$ into two parts
\be
K   \gl K_0\pl U_I(\vec{r}\,) \;,\qquad
K_0 \gl \( -\del^2 \pl m_0^2 \) \;, \qquad
U_I(\vec{r}\,) \gl U(\vec{r}\,)\mi m_0^2 \quad.
\ee
$K_0$ is $\vec{r}$-independent
while the interaction part $U_I(\vec{r}\,)$
depends on the background field.
$~m_0$ is arbitrary and not a physical mass.
But it should be of
the order of the typical mass scale of the problem to
improve the convergence of the series we are generating.

One gets
\bea
\Gamma &=& \frac{1}{2}\, \Tr \ln K_0 \( 1 + {K_0}^{-1} U_I \) \\
 &=& \frac{1}{2}\, \Tr \ln K_0 \;+\;
      \frac{1}{2}\, \Tr \ln \( 1 + {K_0}^{-1} U_I \) \\
 &=& \frac{1}{2}\, \Tr \ln K_0 \;-\;
      \frac{1}{2}\, \Tr \sum_{n=1}^\infty \frac{(-1)^n}{n}
\( {K_0}^{-1} U_I \)^n \quad. \label{expandlog}
\eea
The first term of the right hand side of \eq{expandlog}
does not give any contribution to
the interface tension.
The second term can be evaluated in coordinate
space
\be\label{ml}
\frac{1}{2}\,\sum_{n=1}^\infty \frac{(-1)^n}{n} \int\!dr_1 \ldots \int\!dr_n\;
\,U_I(\vec{r}_1)\, P(\vec{r}_1,\vec{r}_2,m_0) \;\ldots\; U_I(\vec{r}_n)\,
P(\vec{r}_n,\vec{r}_1,m_0)  \quad.
\ee

The main advantage of the splitting of $K$ is
that the `propagator' ${K_0}^{-1}$ can be calculated exactly
\be\label{prop3}
P(\vec{r}_i,\vec{r}_j,m_0) \gl \frac{1}{4\pi |\vec{r}_i-\vec{r}_j|}
\exp\(-m_0 |\vec{r}_i-\vec{r}_j|\)  \quad.
\ee
This is not possible for an arbitrary $\vec{r}$-dependent mass.

Equation (\ref{ml}) contains all corrections to the kinetic terms
included in $\Gamma $ \eqn{Gamma}.
It also gives contributions to the effective potential which can be
removed if one uses the connection of the derivative expansion to the
multi local expansion (cf.~section \ref{CDMLHKE}).
The remaining part, which we are calculating, is
independent of $m_0$ if summed to all orders,
while the volume part gets a $m_0$-dependent
contribution
from the first term of the right hand side of eq.~(\ref{expandlog}).
Truncating the sum in eq.~(\ref{ml}) at finite order $N$ results
of course in
a $m_0$-dependent interface contribution
denoted by $\Gamma_N(m_0)$.
Looking for the $m_0$-independent range
gives a good criterion for the convergence of the series.

The multi local expansion can be adjusted to the interface \eqn{interface}
which depends only on one of three spatial coordinates. Fourier transforming
$~P(\vec{r}_i,\vec{r}_j,m_0)~$ of \eq{prop3} with respect to $x$ and $y$
one gets
\be
{\tilde P}(z_i,z_j,m_0, q) \;=\;
\frac{1}{2\sqrt{m_0^2+q^2}} \; \exp\(- \sqrt{m_0^2+q^2} |z_i-z_j| \) , \qquad
q^2 = p_x^2+p_y^2
\ee
The $q$-integral in each term of \eq{ml} can
now be performed analytically.
The remaining integrals over the $z_i$'s have to be calculated numerically.
They are convergent for $~m_0>0$.

\begin{figure}[t]
\begin{picture}(14.0,7.5)
\put(-0.2,-0.5){\epsfxsize15.0cm \epsffile{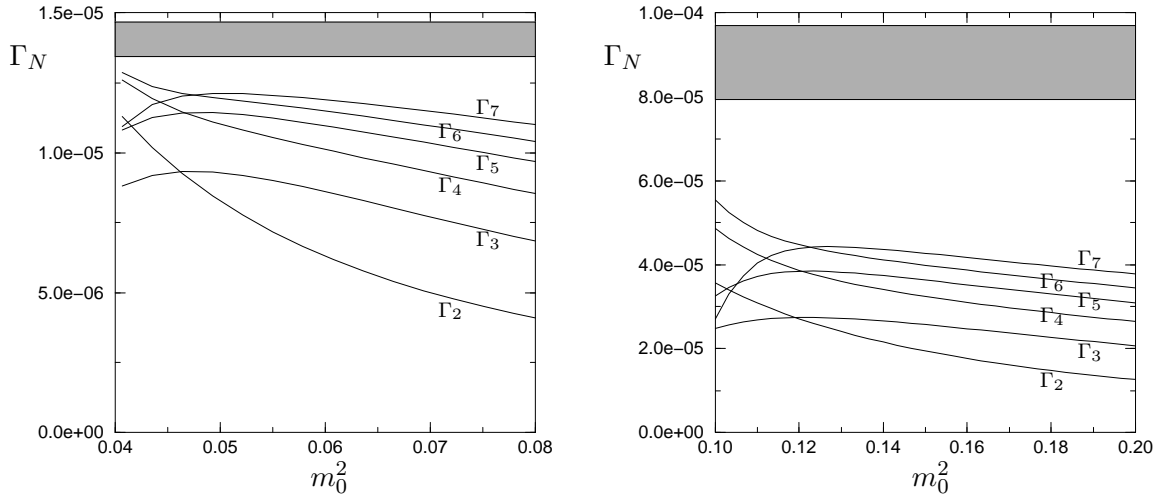}}
\put(0.0,5.6){$\Gamma_N$}
\put(5.7,2.25){\scriptsize $\Gamma_2$}
\put(6.2,3.19){\scriptsize $\Gamma_3$}
\put(5.7,3.9){\scriptsize $\Gamma_4$}
\put(6.2,4.2){\scriptsize $\Gamma_5$}
\put(5.7,4.63){\scriptsize $\Gamma_6$}
\put(6.2,4.95){\scriptsize $\Gamma_7$}
\put(7.9,5.6){$\Gamma_N$}
\put(13.7,1.31){\scriptsize $\Gamma_2$}
\put(14.2,1.7){\scriptsize $\Gamma_3$}
\put(13.7,2.18){\scriptsize $\Gamma_4$}
\put(14.2,2.38){\scriptsize $\Gamma_5$}
\put(13.7,2.63){\scriptsize $\Gamma_6$}
\put(14.2,2.93){\scriptsize $\Gamma_7$}
\put(4.0,0.0){$m_0^2$}
\put(12.0,0.0){$m_0^2$}
\end{picture}
\caption{$\Gamma_N(m_0)$ vs.~$m_0^2$\,. Left plot: the Higgs-fluctuation,
right plot: the $W$-fluctuation. The limits of extrapolation are
expected to be in the hatched range.
The parameters are:
$~\lambda_T/g^2 = 0.024,\; d = 18.3$ corresponding to
$\bar{m}_{\rm H}= 35$ GeV.}
\label{ml_vs_m0-plot}
\end{figure}

Similar to the derivative expansion one can distinguish
between different contributions to the interface tension.
There are those which originate from single Higgs, Goldstone,
$W$, or ghost-loops,
and those which correspond to mixed $W\!$-Goldstone loops.
The ghost-loop cancels 2/3 of the $W$-loop contribution.
The $W$-loops do not have an intrinsic mass scale since
$m_W\ra 0$ in the symmetric phase. The others are massive all over the
interface.

Fig.~\ref{ml_vs_m0-plot}
shows typical results of our calculations for $\laT/g^2=0.024$
and $d=18.3$ which corresponds to a zero temperature Higgs mass of
about 35 GeV. Fig.~1.a shows $\Gamma_N(m_0)$ ($N=2\ldots 7$)
for a typical massive loop (the Higgs mode)
while fig.~1.b shows the same quantity for the $W$-mode.
It turns out that the series behaves badly for small $m_0$.
If $m_0$ is similar to the mass in the middle of the interface or larger
one gets good convergence for the Higgs, the Goldstone, and the mixed
$W$-Goldstone loop.
The   $W$ and ghost loop contributions even in this
range converge very slowly.

\begin{figure}[t]
\begin{picture}(14.0,7.5)
\put(-0.2,-0.5){\epsfxsize15.0cm \epsffile{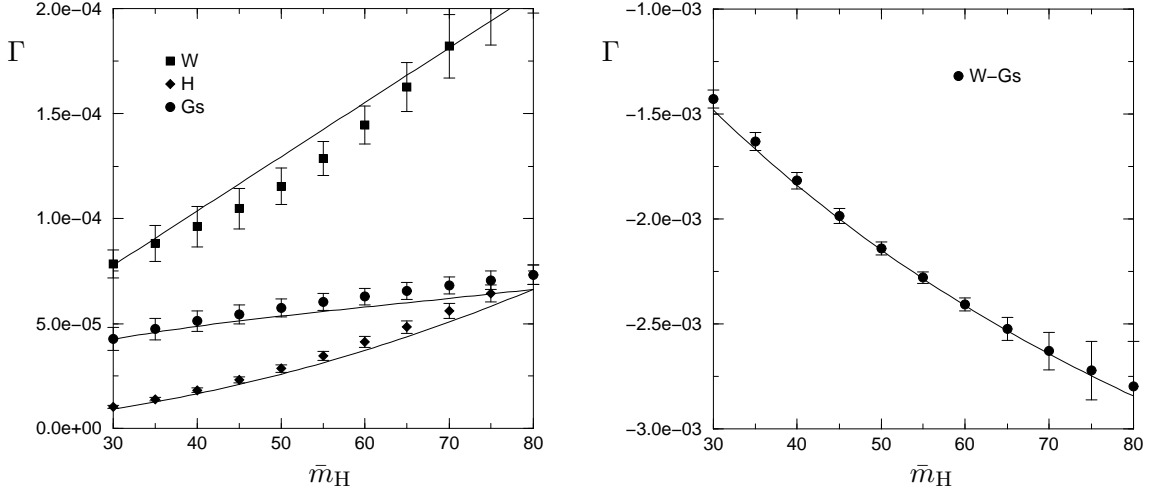}}
\put(0.0,5.6){$\Gamma$}
\put(7.9,5.6){$\Gamma$}
\put(4,0.0){$\bar{m}_{\rm H}$}
\put(12,0.0){$\bar{m}_{\rm H}$}
\end{picture}
\caption{Extrapolated results of the multi local expansion
and the $Z$-factor predictions vs.~$\bar{m}_{\rm H}$
(eq.~\protect\ref{mH}).
The full lines are the $Z$-factor predictions.}
\label{ml-results-plot}
\end{figure}

We extrapolated $\Gamma_N(m_0)$ to $N=\infty$ for several values of $m_0$.
{}From the
$m_0$-dependence of the results we estimated an error of the limit.
In fig.~\ref{ml-results-plot} the limits with error bars of our
extrapolations are plotted
for several
values of $\bar{m}_{\rm H}\,$ \eqn{mH}.
Fig.~\ref{ml-results-plot}.a shows
the single Higgs, Goldstone and
$W$-contributions (single modes)
while fig.~\ref{ml-results-plot}.b shows
the mixed $W\!$-Gs-loop.
The full lines are the predictions from the $Z$-factor for the corresponding
quantities.

One finds small deviations for the single modes (less than 10\%).
The $W$-Gs-con\-tri\-bu\-tion agrees within the errors with the
$Z$-factor prediction.
The large errors are due to the extrapolation to $N\!=\!\infty$.
At any finite order the deviations are smaller as we will
show in section \ref{CDMLHKE}.
The $W$-Gs-con\-tri\-bu\-tion
is an order of magnitude larger than the sum of the single modes
contributions. This depends of course on
the gauge-fixing parameter $\xi$
which is 1 in our case.

\subsection{The Heat Kernel Expansion}\label{HKE}

Another way to calculate the higher derivative terms
of the one-loop effective action is the
heat kernel expansion \cite{SeDeW}. $\Gamma$ \eqn{Gamma}
is expressed as
\be\label{hk1}
\Gamma \gl
\frac{1}{2} \int_0^\infty\!\frac{dt}{t} \exp\(-m_0^2 t\) \Tr
        \exp\( -t (-\del^2 + U_I) \)
\ee
$m_0$ and $U_I(\vec{r}\,)$ are defined in the same way as above.
$\Tr\exp\( -t (-\del^2 + U_I) \)$ is now expanded in powers of $t$.
Using world-line methods this can be done in a very economical
way. One finally gets
\be
\Gamma \gl
\frac{1}{2} \int_0^\infty\!\frac{dt}{t} \(4\pi t\)^{-d/2} \exp\(-m_0^2 t\)
\sum_{n=1}^\infty (-t)^n O_n
\ee
where the $O_n$'s are functionals of the field $\ph(\vec{r}\,)$.
In reference \cite{FliegnerEA2} they are given up to $O_8$;
with the new world-line method they can easily be calculated to even
higher order.
For $~n\!>\!d/2~$ the $t$-integral gives $~\Gamma(n\!-\!d/2)$.
The dimension $d$
is in our case 3. Using this formula for  $~n\!\le\! d/2~$ as well
defines a dimensional regularization of the ultraviolet divergences.

Every order of this expansion is again free of infrared divergences
for $~m_0\!>\!0$.
Trun\-ca\-ting the sum at finite order $N$ one generates $m_0$-dependent
terms although the limit is independent of $m_0$. This aspect
is very similar
to the multi local expansion. Indeed even the numerical results are
nearly the same if one compares the $N$th order of the
multi local expansion with the ($N$+1)th order of the heat kernel expansion,
though the higher derivatives are truncated in the heat kernel
approach.
One can again extrapolate to infinite order. The results agree very well
with those of the multi local expansion.
This shows that at low order of the corresponding
expansion higher order contributions in 1/$d$  do
not play any significant role at small $\laT$ (large $d$).
Possible large expansion parameters do not arise.
This need not be the case at higher orders, as
discussed in chapter \ref{MM}.

\subsection{Connection between Derivative,  Multi Local, and
            Heat Kernel Expansion} \label{CDMLHKE}

The good agreement between the limit of large $N$ of
the multi local expansion and of the heat
kernel expansion confirms that our results are reliable.
The agreement is already very good at small $N$ and this indicates
that the higher derivative terms are numerically very small.
Indeed the large-$N$ limit is very well approximated
by the $Z$-factor prediction.
In order to separate the higher derivative terms in the two expansions
it is necessary to know how the effective potential and the $Z$-factor
build up.

The operators $O_n$ of the heat kernel expansion are already local.
One can sort directly for powers of derivatives.
The terms without explicit derivatives are of the form
\be\label{noDer}
\frac{1}{2}\int_0^\infty\!\frac{dt}{t} \(4\pi t\)^{-d/2} \exp\(-m_0^2 t\)
\sum_{n=1}^\infty (-t)^n \frac{1}{n!} \int d^3r \; U_I(\vec{r}\,)^n
\ee
If one has a 1$\times$1-matrix
$~U_I=\(m(\vec{r}\,)^2-m_0^2\)~$ these terms sum up
to give the $m(\vec{r}\,)^3$-term
of the effective potential (cf.~\eq{Veff}).

\begin{figure}[t]
\begin{picture}(14.0,7.5)
\put(8.0,0.0){$\exp(-z/d)$}
\put(1.0,-0.5){\epsfxsize13cm \epsffile{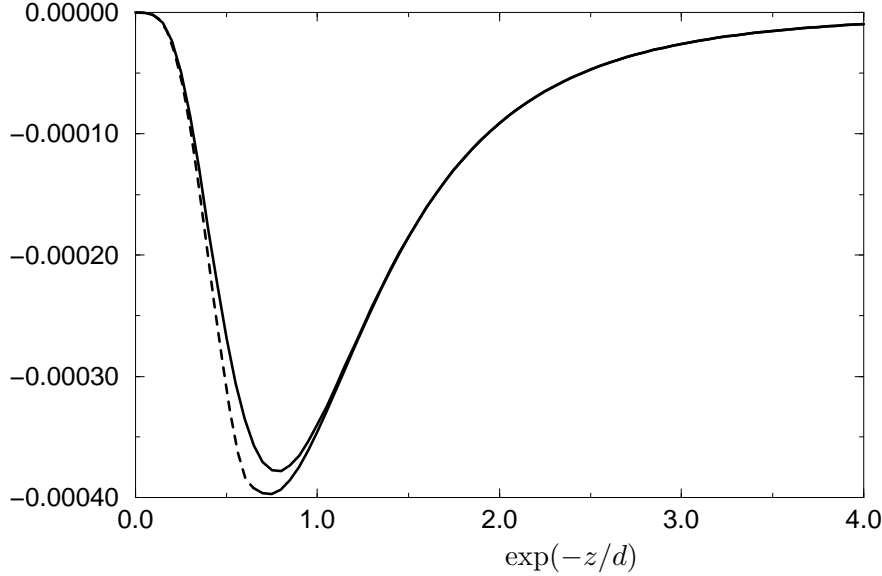}}
\end{picture}
\caption{The integrand of eq.~(\protect\ref{evint}) and the $Z$-factor
prediction vs.~$\exp(-z/d)$.
The parameters are $\lambda_T/g^2=0.04, d=12.5$}
\label{matrixproblem}
\end{figure}

If $U_I$ is the $2\!\times\!2$-matrix
$~\(U_2-m_0^2\)~$, with $U_2$ of \eq{M2},
even this term contains derivatives. The two-derivative order of this
term gives the mixed $W\!$-Gs-loop contribution to $Z_{\rm H}$ \eqn{Zfkt}.
$U_2$ can be diagonalized to give
\be\label{Kd}
\left(
  \begin{array}{c}
    \frac{1}{2}\mWq+\frac{1}{2}\mGsq+
    \frac{1}{2}\sqrt{\mWf-2\,\mWq\mGsq+\mGsf+4\,(\del_z\ph)^2}
      \dquad 0\dquad\\[1ex]
     \dquad 0 \dquad
    \frac{1}{2}\mWq+\frac{1}{2}\mGsq -
    \frac{1}{2}\sqrt{\mWf-2\,\mWq\mGsq+\mGsf+4\,(\del_z\ph)^2}\\
   \end{array}
\right)
\ee
In the background of the interface \eqn{interface} one eigenvalue
$\lambda_+$ is positive over the whole $\ph$-range
while the second one $\lambda_-$ turns
negative for small $\ph$. Summing up all orders of $n$ one gets
\be
 \int\! d^3r \(     \lambda_+(\vec{r}\,)^{3/2}
                \pl \lambda_-(\vec{r}\,)^{3/2} \) \quad.
\ee
The contribution to the surface tension is proportional to
the $z$-integral of
\be\label{evint}
-3 \( \lambda_+^{3/2} \pl \lambda_-^{3/2}
       \mi \mGs^3 \mi \mW^3\)
\ee
In fig.~\ref{matrixproblem} the integrand is plotted in comparison
to the $Z$-factor prediction for this contribution.
In the region where
$\lambda_-$ is negative the real part has been taken.
This is shown by the dashed line.
One sees that the  $Z$-factor prediction
is a very good approximation over the whole $z$
(respectively $\ph$) range.

The terms in the heat kernel expansion which have two explicit
derivatives are of the form
\be\label{twoDer}
\frac{1}{24}\;\int_0^\infty\!\frac{dt}{t} \(4\pi t\)^{-d/2} \exp\(-m_0^2 t\)
(-t)^3  \int d^3r \; \(\del U\)^2 \, \frac{1}{n!}\,(-U_I\,t)^n
\ee
For a single mode they sum up to the $Z$-factor contributions.
Nevertheless, this series converges only slowly
in the massless channel.
Since the higher derivative terms are numerically small
this is the main reason for the bad convergence of the
heat kernel expansion for the massless mode. The same  applies
to the multi local expansion
(cf.~fig.~\ref{ml_vs_m0-plot}.b).

If one is interested only in the higher derivative terms one has to
subtract those terms which contribute to the $Z$-factor.
The remaining contribution to the interface tension
turns out to be very small ($\lsim$ 1\%) compared to the
$Z$-factor contribution to all orders we calculated.
This again suggests that
it can be neglected in good accuracy.

The connection between the multi local and the derivative expansion
is established by expanding every order locally.
If one substitutes in \eq{ml} the integration variables via
\bea
\vec{R} \!\!&=&\!\! \frac{1}{n} \( \vec{r}_1 + \ldots + \vec{r}_n \)\\
\vec{\rho}_1 \!\!&=&\!\! \vec{r}_2-\vec{r}_1 \,,
\quad \ldots\,,\quad \vec{\rho}_{n-1}
\!\!\gl\!\! \vec{r}_n-\vec{r}_{n-1}
\eea
and Taylor-expands $U_I(\vec{r}_i)$ around $\vec{R}$ the
$\vec{\rho}_i$-integrals can
be done analytically. Summing up all terms without derivatives
one gets the effective potential.
We used this connection to separate the interface from the volume
contributions.

Summing up all terms with two derivatives regenerates the $Z$-factor
contribution. The multi local expansion is an expansion in powers of
fields. Therefore this contribution builds up by terms which contain
two derivatives and 2$N$ fields.
Comparing with \eq{twoDer} one confirms that
these must be the same terms which connect the heat kernel expansion
with the derivative expansion.
Hence the  $Z$-factor contribution is generated in both expansions
in the same way.
The $N$th order of the multi local expansion corresponds to
the $(N\!+\!1)$th order of the heat kernel expansion.
If the higher derivative
terms are small one expects the corresponding orders of both expansions
to give similar results. This behavior was indeed observed by us.

In conclusion of this chapter both methods,
the multi local and the heat kernel expansion, although
quite different from the calculational point of view,
yield the same results
because the higher derivative terms are very small
at the order of the expansion that has been studied.
For interfaces typical for
zero temperature Higgs masses up to 80 GeV the
$Z$-factor of the derivative expansion gives numerically
nearly all corrections to the kinetic term of the action.
The massless mode ($W$-boson) would
need higher order treatment in these expansions.
The additional contribution is expected to be numerically unimportant.
In the next chapter we investigate this mode in more detail.

\section{The Massless Mode}\label{MM}

\subsection{Resummation of Infrared Divergent Heat Kernel Operators}
\label{RIDHKO}

For ``small'' Higgs mass $\mH\le\mW~$, i.e.~small $\laT/g^2~$,
and $g_3^2$ the wall thickness $d$ (in units of $(gv_0(T))^{-1}$)
is large (cf.~\eq{d}).
This suggests a large-$d$ expansion of the one-loop
contributions to $\sigma(d)$. The appropriate tool would to be
the derivative expansion
\be\label{dexpansion}
\sigma(d) \gl c_{\del^0}\,d \;+\; \frac{c_{\del^2}}{d}
 \;+\; \frac{c_{\del^4}}{d^3}  \;+\; \calO\!\(\frac{1}{d^5}\)
\ee
where the first term corresponds to the potential, the next one
to the wave function contribution, and the following to
the four derivative terms of the effective action.
This expansion, however, breaks down due to infrared
singularities coming from the massless degrees of freedom
as we have shown above.
Already the $c_{\del^4}$-term does not exist.

Obviously the question comes up whether one can rearrange
the various contributions in such a way that a meaningful
large-$d$ expansion arises. It will turn out that the
first two terms in \eq{dexpansion} are not modified
and remain leading. The next term (in the massless sector)
will be of order $d^{-2}$
not present in the derivative expansion.
In order to see this behavior and to isolate this contribution
it is useful to sum up ``towers'' of related terms in the
heat kernel expansion ``vertically''.
This  means that one does not evaluate the complete heat kernel
coefficient  but only parts.
These parts
have the same number of derivatives and are labeled with $\alpha$.
If one does the summation without caring about divergences
one regenerates the derivative expansion but this is not
correct in the infrared region. Here it is better to
use the exponential falloff of the background to rearrange
the series.

Setting $m_0\!=\!0$ in \eq{twoDer} one sees that
the  one-loop wave function contribution to the interface tension is given by
a tower of terms of the form
$~ t^n \, U(z)^n \, (\del U(z))^2 ~$.
These are generated by an exponential $~\exp(-U t) (\del U)^2~$
\be
\sigma_{\del^2} \gl
\frac{c_{\del^2}}{d} \gl
\frac{1}{24}\; \int_0^\infty \frac{dt}{t} \,
(4\pi t)^{-3/2} \, t^3 \;
\int dz (\del U)^2\exp(-U t) \quad.
\ee
At higher orders of $t^n$
we obtain the general structure
\be\label{HKsum1}
\sigma_{\alpha,\beta} \gl
\frac{1}{2} \int_0^\infty \frac{dt}{t} \,
(4\pi t)^{-3/2} \, t^{\beta+\alpha/2}
\int dz \; O_{\alpha\beta}\exp(-U t)
\ee
where the first index ($\alpha\ge2$) is the number of derivatives
and the second one ($0\le\beta\le\alpha$) labels the towers summed over;
$\beta$ is equal to the power of $U$'s.
A similar rearrangement of the heat kernel operators was already obtained
in ref.~\cite{FliegnerEA1}.
Up to order six the operators $O_{\alpha,\beta}$ are given by
\begin{eqnarray*}
O_{2,2} &=& \frac{1}{12}\; \(\del U\)^2 \\
O_{4,2} &=&  -\; \frac{1}{120}\; \(\del\del U\)^2 \\
O_{4,3} &=& \frac{1}{72}\; \del\del U \(\del U\)^2 \\
O_{4,4} &=& -\;\frac{105}{30240}\; \(\del U\)^4\\
O_{6,2} &=& \frac{1}{1680}\; \(\del\del\del U\)^2 \\
O_{6,3} &=& -\;\frac{1}{30240}\; \(22\,(\del\del U)^3 \;+\;
                              84\,\del\del\del U\, \del\del U\, \del U\) \\
O_{6,4} &=& \frac{1}{9979200}\; \(30030\,(\del\del U)^2 (\del U)^2  \;+\;
                              7700\,(\del\del\del U)(\del U)^3\) \\
O_{6,5} &=& -\; \frac{300300}{259459200}\; \((\del\del U)(\del U)^4 \) \\
O_{6,6} &=& \frac{1051050}{10897286400}\; \(\del U\)^6
\end{eqnarray*}

Using \eq{HKsum1} for the massless $W$ and ghost modes
the $t$-integrals are infrared
divergent, i.e.\ for $t\rightarrow \infty$,
because of the behavior of $U(z)=m_W(z)^2=\ph(z)^2/4$
at $z\rightarrow\infty$.
With the ansatz \eq{interface} $U$ behaves as
\bea\label{approxK}
U(z\rightarrow\infty) &=& \frac{1}{4}\exp\(-4 z/d\)\\
\del U(z\rightarrow\infty) &=& -\frac{4}{d}\, U(z) \quad.
\eea
Using this interface shape we can now
introduce $U$ as integration variable
\be
\frac{dz}{d} \gl -\frac{1}{4}\frac{dU}{U}
\ee
and rewrite \eq{HKsum1} as
\bea\label{HKsum2}
\sigma_{\alpha,\beta} &=&
c_{\alpha,\beta}\; \int_0^\infty \frac{dt}{t} \,
(4\pi t)^{-3/2} \, t^{\beta+\alpha/2}
\(\frac{-4}{d}\)^{\alpha-1}
\int_0^\epsilon \frac{dU}{U} U^\beta \exp(-U t)\\
&=&
c_{\alpha,\beta}\; \int_0^\infty \frac{dt}{t} \,
(4\pi t)^{-3/2} \, \(\frac{t}{d^2}\)^{\alpha/2} d
\(-4\)^{\alpha-1} f(\beta) \quad.\label{HKsum3}
\eea
$f(\beta)$ is the $U$-integral rescaled with $t$.
\be
f(\beta) \gl
\int_0^{\epsilon t} \frac{du}{u} u^{-\beta} \exp(-u) \;\;
\begin{array}[t]{c}
\mbox{\Large $\longrightarrow$}\\[-0.5ex]
\mbox{\scriptsize $d\!\rightarrow\!\infty$}\\[-1ex]
\mbox{\scriptsize $t/d^2$~fixed}
\end{array}
\;\; \Gamma(\beta)
\ee
The upper integration limit
$\epsilon$ must be small enough for the approximation
\eq{approxK} to be valid. The remaining part of the
$z$-integration in \eq{HKsum1} is not related to
the infrared part and needs no resummation.
The two derivative term $\alpha=2$ is finite and can be
treated separately as well. It generates the one-loop contributions
to the $Z$-factor.

The $\alpha\ge3$ terms in \eq{HKsum3} diverge at large $t$. All these
terms have to be resummed.
Rescaling $~\tilde{t}=t/d^2~$ we get
\be
\sigma_{\alpha,\beta} \;\;\cong\;\;
c_{\alpha,\beta}\;
\(-4\)^{\alpha-1} \Gamma(\beta) \;\frac{1}{d^2}\;
 \int_0^\infty \frac{d\tilde{t}}{\tilde{t}} \,
(4\pi \tilde{t}\,)^{-3/2} \, \tilde{t}^{\alpha/2}
\quad.\label{HKsum4}
\ee
The $c_{\alpha,\beta}$'s are
\bc
\bt{|rr|r|r|}
\hline
$\alpha$ & $\beta$ & $c_{\alpha,\beta} \Gamma(\beta)$ &
 $c_\alpha = \sum_\beta c_{\alpha,\beta} \Gamma(\beta)$ \\
\hline
4 & 2 & -2.133 &  \\
 & 3 & 7.111 &  \\
 & 4 & -5.333 & -0.356\\[1ex]
\hline
6 & 2 & 2.438 & \\
 & 3 & -28.715 & \\
 & 4 & 92.919 & \\
 & 5 & -113.718 & \\
 & 6 & 47.497 & 0.271\\
\hline
\et
\ec
\vspace{0.5ex}
This clearly shows strong cancelations between towers of operators
starting in different order of the heat kernel expansion.
This should have a deeper reason which we still do not understand.

The contribution from the higher derivative terms is summarized by
\be\label{HKsum5}
\sum_{\alpha\ge 4,\,\beta } \sigma_{\alpha,\beta}  \;\;\cong\;\;
\frac{1}{d^2} \int_0^\infty  \!
\frac{d\tilde{t}}{\tilde{t}} \; \frac{1}{(4\pi\tilde{t})^{3/2}} \;
\sum_{\alpha\ge4} (-4)^{\alpha-1}\; c_\alpha\; \tilde{t}^{\alpha/2}
\quad.
\ee
This equation is of course only a formal expression
since the $\tilde{t}$-integrals do not exist in the infrared
($\tilde{t}\!\rightarrow\!\infty$) term by term.
It is likely, however,
that the complete 1-loop contribution has a $1/d$ expansion,
perhaps in the sense of an asymptotic series. This would
correspond to a resummation of the series
$~\sum_{\alpha\ge4} (-4)^{\alpha-1} \;c_\alpha\;
\tilde{t}^{\alpha/2}~$
in such a way that the $\tilde{t}$-integral exists.
We did not attempt to prove this assertion
because the heat kernel
expansion is not an appropriate tool for this purpose.
Determining $c_\alpha$ involves heat kernel contributions
up to order $3\alpha/2$. We saw dramatic cancelations
between the different heat kernel contributions of
different order to the same $c_\alpha$.
A more appropriate framework for calculating the coefficient
of the $1/d^2$ term might be the treatment of the eigenvalue
spectrum of the equivalent quantum mechanical problem
with the Hamiltonian $-\del_z^2 +\frac{1}{4}\ph(z)^2$.

The $1/d^2$ contribution arises from the
region $~t\!\sim\! d^2~$ and can not be seen in
the multi local or the heat kernel expansion in finite order
because these expansions cover only a finite $t$ range
(in units of $v$).
The natural order of magnitude for this non-leading
correction to the predicted $Z$-factor contribution
of the massless mode is therefore 5-10\% at intermediate
Higgs mass (assuming a coefficient of order one).

\subsection{Derivative Expansion with a Non-Perturbative Infrared Cutoff}
\label{IPDE}

Because the infrared region is sensitive to non-perturbative
effects (confinement) one could also take the point of view of
cutting-off this contribution by introducing a magnetic mass
$~m_\gamma=\gamma g^2 T\,$.
Strictly speaking non-perturbative
effects also change the vertex functions.
Introducing a cut-off mass is therefore at best an approximate
tool.
It would enter in
\eq{HKsum4} as a factor $~\exp(-m_\gamma^2 d^2 \tilde{t})~$.
The exponent can easily be estimated using \eq{d} and
$~g_3(T)=gT/v(T)~$
\be\label{mgamma}
m_T^2 d^2 \gl
\gamma^2 g^4 T^2 \frac{8}{\laT/g^2} (gv(T))^{-2} \gl
\gamma^2  \(\frac{gT}{v(t)}\)^2 \frac{8}{\laT/g^2} \gl
\gamma^2 64\pi\,g_3^2 \quad.
\ee
This is not small for any interesting Higgs mass, if
$~\gamma\gsim 1/(3\pi)~$ (cf.~ref.~\cite{BuFoHeWa}).
Therefore
possible large $\tilde{t}$, i.e.~infrared,  contributions to \eq{HKsum4}
are strongly suppressed if one introduces such a cut off mass $m_\gamma$.
If such contributions would become big in perturbation theory
this would be a typical infrared effect and not be physical, because
perturbation theory is not applicable in this range.
However, we do not expect that.

Since the higher derivative terms turn out to be small
and therefore the effective action apparently is well described
by the potential and the $\del\ph\del\ph$ term it might be
useful to look to the derivative expansion again.

There are three scales in the problem. \\
(i)  The inverse wall thickness
\be
\frac{1}{d} \gl \sqrt{\frac{\laT}{8 g^2}}\; g\,v(T)
\ee
derivatives of the interface have this dimension.\\
(ii) The $\vec{r}$-dependent field
\be
g\, \frac{\Phi(\vec{r}\,)}{\sqrt{2}} \gl g\, v(T)\, \ph(\vec{r}\,) \quad.
\ee
($\ph(\vec{r}\,)$ is normalized to 1 in the broken phase)\\
(iii) A genuine tree mass $m_3$ in the Higgs and Goldstone mode
and an infrared scale $m_\gamma$ in the massless modes, respectively.
We will denote both masses with $m$ in
this section.

The derivative expansion is valid if
\be\label{ab}
a) \qquad \frac{1}{d} \kl m
\qquad\qquad {\rm or} \qquad\qquad
b) \qquad  \frac{1}{d} \kl g\, \frac{\Phi(\vec{r}\,)}{\sqrt{2}}
\ee
since a typical behavior of a tower with $\alpha$ derivatives
starting at power $\beta$ of $\Phi^2$ is
\be
\(\frac{1}{d}\)^\alpha \Bigg/
\(\frac{ g^2 \Phi( \vec{r}\,)^2}{4} +m^2 \)^{\alpha/2+\beta-3/2}
\quad.
\ee
$a$) means $~d^2 m_3^2>1~$ or at
the critical temperature in the massless channel
$~d^2 m_\gamma^2 = \gamma^2 g_3^2\, 64 \pi > 1~$ \eqn{mgamma}.
Given some (unknown) value of $\gamma$ ($\gamma=\calO(1)/(3\pi)$ in the
work of ref.~\cite{BuFoHeWa}) this is not fulfilled for correspondingly
small $g_3(T)^2$ (i.e.~$\laT$ small).\\
$b$) means
\be
\ph \gr \frac{\sqrt{\laT/g^2}}{\sqrt{8}} \Sim \frac{m_{\rm H}}{8\, m_W}
\qquad {\rm at~tree~level.}
\ee

Thus for large $\mH > \mW$, i.e.~$g_3^2$ on the upper end limit
of perturbation
theory, small magnetic masses can be specified because ($a$) is fulfilled.
For small $\mH$ the range of allowed $\ph$'s extends almost to $~\ph=0~$
with condition ($b$), though condition ($a$) might not be fulfilled
anymore.

The inequalities ($a$) and ($b$) in (\ref{ab})
have also been checked numerically
and turn out to be quite accurate without any further factors.

In the multi local expansion $1/(md)$ has not to be small since all
powers of the derivatives are summed but one still needs an
expansion in powers of $(\ph^2-m_0^2)/m^2$. Since we already know
that higher derivative terms are small and that the $\del\ph\del\ph$
term dominates we can see very explicitly that this series converges very
slowly for the massless mode.

\begin{figure}[t]
\begin{picture}(14.0,7.5)
\put(7.0,0.0){$m_\gamma^2/(gv(t))^2$}
\put(0.3,6.5){\mbox{\Large $\frac{\Gamma_{\del^n}}{(gv(T))^2}$}}
\put(1.0,-0.5){\epsfxsize13cm \epsffile{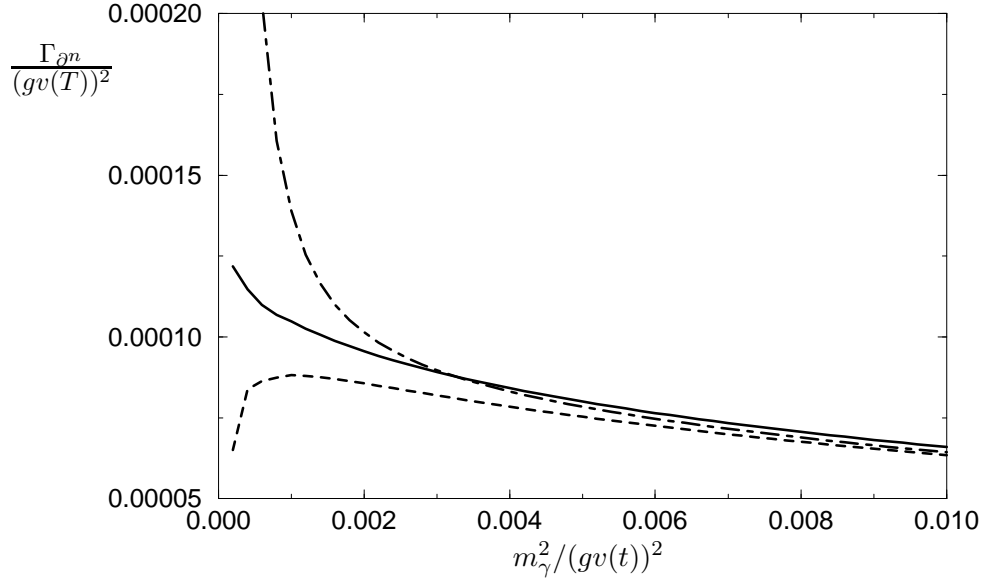}}
\end{picture}
\caption{The one-loop contribution
to the surface tension in the derivative expansion.
The full line is the $Z$-factor contribution; the
dashed line includes the terms up to order $\del^4$;
the dot-dashed line up to $\del^6$.}
\label{deplot}
\end{figure}

Having introduced a cutoff mass the derivative expansion can
be studied term by term because now the  $z$-integrals
no longer diverge.
In fig.~\ref{deplot}
we display the
$Z$-factor contribution and the results including
the next two orders of the
derivative expansion
for $\laT/g^2\!=\!0.04$ and \mbox{$d\!=\!12.5$}
vs.~$m_\gamma$.
We find that
a well-behaved converging expansion is obtained
for $~m_\gamma^2\gsim 0.002(gv)^2\,$.
This is in agreement with the inequality obtained from (\ref{ab}.a).
It also implies a cutoff for $\ph\!<\!0.09$.
This is a rather small cutoff, small compared to the dynamic mass scale
set by the bound state masses in the symmetric phase.

The derivative expansion including the $Z$-factor is
therefore a good approximation to the effective action
if a small infrared mass $m_\gamma$ is introduced.
The higher derivative terms are under complete control.
Of course such a $m_\gamma$ influences the $Z$-factor
contribution to the interface tension as well and thus
parameterizes the effects outside the range of perturbation
theory.

We restrict ourselves to the one-loop $Z$-factor in the rest of this paper,
though considering its strong gauge fixing dependence a two-loop
calculation would be highly desirable \cite{Wir5}.

\section{The Interface Tension}\label{IT}

Taking into account the $Z$-factor modifies on the one
hand the critical bubble or the planar interface solutions
and gives on the other hand direct
additional contributions to the effective action of
the bubble surface or the interface.
For definiteness we concentrate on the planar interface.
Thin walled critical bubbles can be treated in the same way.

Starting from the effective action \eqn{Seff} neglecting the
${\cal O}(\partial^4)$-terms the saddle point equation reads
\be
Z_{\rm H}(\varphi)\,\partial^2\varphi
\pl
\frac{1}{2} \frac{\del Z_{\rm H}(\varphi)}{\del \ph}\,
(\partial_i \varphi)^2 \gl \frac{\del V_{\rm eff}(\varphi)}{\del \ph} \quad.
\ee
With the substitution
\be
V_{\rm eff}(\varphi)\gl \tilde V(\varphi)\, Z_{\rm H}(\varphi)
\ee
we obtain
\be\label{mspeq}
Z_{\rm H}(\varphi)\,\partial^2\varphi \gl
Z_{\rm H}(\varphi)  \frac{\del\tilde{V}(\varphi)}{\del \ph}
\pl
\frac{\del Z_{\rm H}(\varphi)}{\del \ph}
\(\tilde{V}(\varphi)\,-\,\frac{1}{2}(\partial_i\varphi)^2 \) \quad.
\ee
The boundary conditions of a planar interface $~\ph(\vec{r}\,)\!=\!\ph(z)~$
at the critical temperature are
\be
\ph(z\!\rightarrow\!-\infty)=\ph_S = 0\, , \qquad
\ph(z\!\rightarrow\!\infty)=\ph_B =1
\ee
where $~\ph_S~$ and  $~\ph_B~$ are the field values of the
symmetric and the broken minimum, respectively, and
$~~V_{\rm eff}(\ph_S)= V_{\rm eff}(\ph_B)=0~$.
Eq.~(\ref{mspeq}) is solved for a general positive $Z_{\rm H}(\ph)$ by
\be
\frac{d\varphi}{dz}=\sqrt{2\tilde{V}(\varphi(z))} \quad.
\ee
The interface tension of this solution is
\be
\sigma \;=\;
\int_{-\infty}^\infty\!\!dz\, \( \frac{1}{2} \;Z_{\rm H} \(\del_z\ph(z)\)^2
\;+\; V_{\rm eff}\(\ph(z)\) \)
\ee
where $V_{\rm eff}(\varphi)$ has to be evaluated at the critical
temperature. Substituting the integration variable one gets an
expression which does not refer to the explicit interface
solution \cite{Wir1}
\be\label{sigma}
\sigma \;=\; \int^{\varphi_B}_{\varphi_S} \!\! d\varphi \;
\sqrt{2\, Z_{\rm H}(\varphi)\, V_{\rm eff}(\varphi)}
\ee

\begin{figure}[t]
\begin{picture}(14.0,7.5)
\put(0.7,6.2){\mbox{\Large $\frac{\sigma}{(g^2T)^3}$}}
\put(7.0,0.1){$\laT/g^2$}
\put(0.5,-0.5){\epsfxsize13cm \epsffile{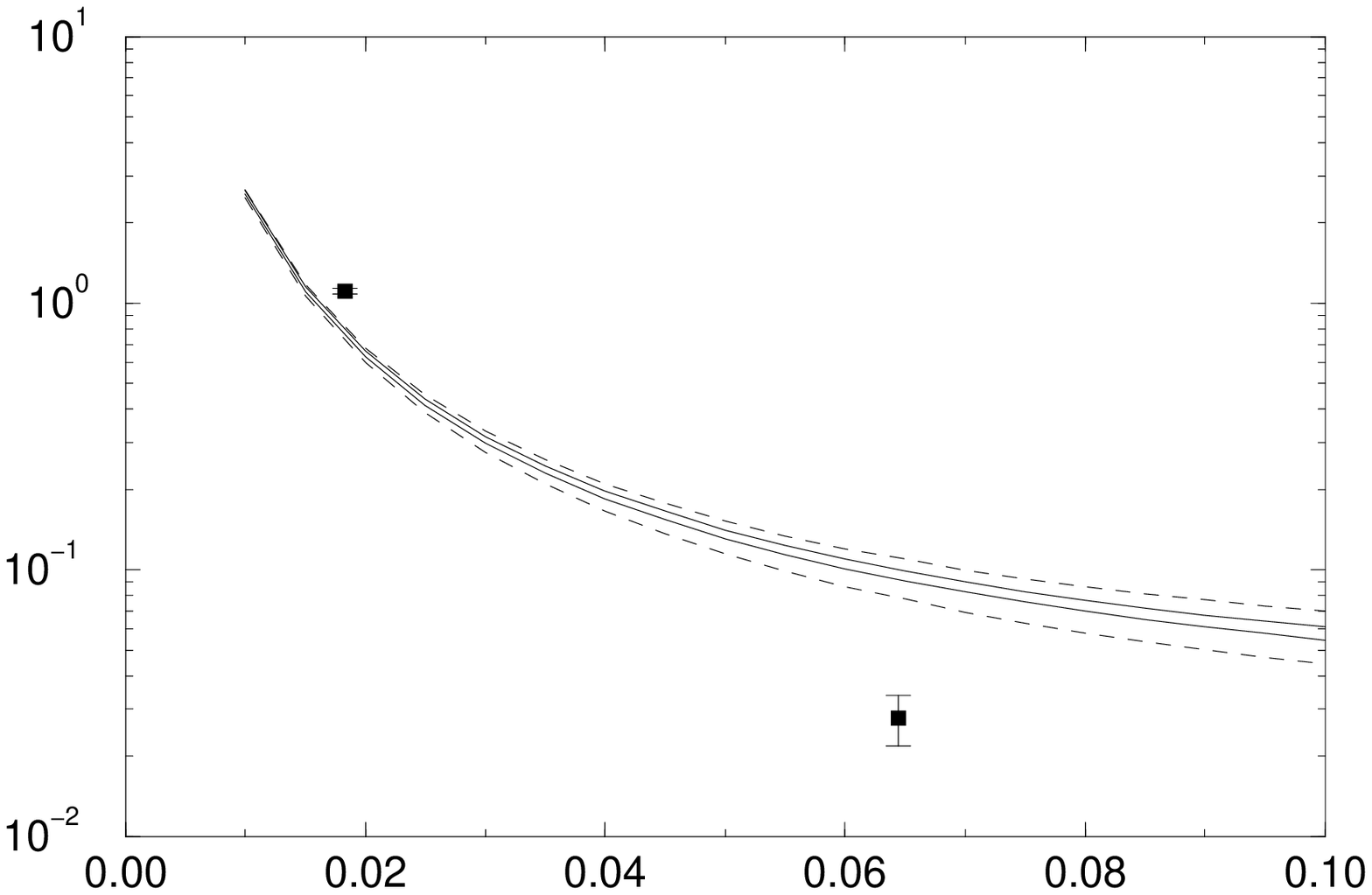}}
\end{picture}
\caption{The interface tension vs.~$\laT/g^2$.
$~\sigma$ of eq.~(\protect\ref{sigma}) is plotted in the upper two lines;
$\tilde{\sigma}$ of eq.~(\protect\ref{sigmalit}) in the lower two lines.
The dashed lines are the
Feynman gauge results, the full lines are the Landau gauge values.
The two data points are the lattice values of
ref.~\protect\cite{KajantieEA2}.}
\label{interfactension}
\end{figure}

$\sigma$ of \eq{sigma} is the generalization
of the $Z\!=\!1$ formula used in the
literature as perturbative value of the interface tension
\be\label{sigmalit}
\tilde{\sigma} \;=\; \int^{\varphi_B}_{\varphi_S} \!\! d\varphi \;
\sqrt{2\, V_{\rm eff}(\varphi)} \quad.
\ee
Eq.~(\ref{sigma}) and (\ref{sigmalit})
are now used to investigate the influence of
the one-loop corrections to the kinetic term on the interface tension.
Using the two-loop effective potential of ref.~\cite{Wir2}
and the one-loop $Z$-factor of \eq{Zfkt} we calculated
$\sigma$ (lower two lines) and $\tilde{\sigma}$ (upper two lines)
for several values of $\lambda_T$
and the gauge fixing parameters $\xi\!=\!0$ (Landau gauge, full lines) and
$\xi\!=\!1$ (Feynman gauge, dashed lines).
In the latter case the real part has been taken where $Z$ becomes
negative.
The results are shown in
fig.~\ref{interfactension}.

In both gauges the interface tension is lowered due to the $Z$-factor
contribution. The dependence on the gauge fixing parameter via
the effective potential is over-compensated.
It will be interesting to see whether this behavior improves
if the two-loop $Z$-factor is used.
The two data points in fig.~\ref{interfactension} are the lattice
values of the interface tension calculated in
ref.~\cite{KajantieEA2}.
The data point of ref.~\cite{CsikorEA} was calculated on a four
dimensional lattice and is difficult to compare.
It is at a Higgs mass of 35 GeV
and roughly agree with the two-loop perturbative
contribution. At large Higgs mass the lattice data
fall significantly below the perturbative result.
The 60 GeV data point of ref.~\cite{KajantieEA2} is
shown in fig.~\ref{interfactension}. The trend towards a  smaller
interface tension is even more pronounced at larger Higgs
masses \cite{GuertlerEA}. One should keep in mind that
these lattice data are plagued with systematical uncertainties
more important than the statistical error.
Nevertheless, these results indicate that a quasiclassical
effective potential approach may become completely unreliable
in a Higgs mass range starting at about 60 GeV.
A magnetic mass of the order $~m_\gamma^2\sim 0.002(gv)^2\,$
(cf.~fig.~\ref{deplot}) would not be sufficient to
explain such effects.
In view of the good agreement of other quantities, e.g.~the
latent heat, this may appear surprising. One should keep in
mind, however, that the interface tension is the only
quantity out of the measured ones really testing the shape
of the effective potential.

\section{Conclusion}\label{Conclusion}

We started with the observation that the $Z$-factor is very gauge dependent
and rather infrared sensitive. The derivative expansion for integrated
quantities like the interface tension
breaks down at the $\calO(\del^4)$
order due to massless modes obtained if one continues the perturbative
description to small $\ph$-values. One can treat the technical problem
how to resum the higher derivative terms and we argued that
there is a finite contribution of the massless channel to the interface
tension.

Our numerical evaluations strongly indicate that this contribution to the
higher derivative terms is small.
In praxi the order of the multi local and
the heat kernel expansion is limited
and this kept us from proving this with our methods.
A large contribution,
if it arises nevertheless, would be a pure infrared effect.
It would be based on a $\ph$-range which is perturbatively not
accessible and hence be an artifact which had to be substituted by
a nonperturbative calculation in the hot phase.
Introducing a infrared cut-off mass one suppresses such a contribution.

For small zero-temperature Higgs-masses ($\lsim$ 70 GeV) perturbation
theory works well in the broken phase.
The infrared problems which plague the higher order derivative terms
are restricted to even smaller $\ph$-values ($<$ 0.1$gv(T)$).
The massive modes are free of infrared divergences in the whole
$\ph$-range.
Our numerical calculations showed
that the higher derivative terms
give only very small contributions to the interface tension
for these modes. They add a maximum of some percent in any order
we calculated to the leading $Z$-factor contribution.
In derivative expansion they are suppressed by $1/d^2$
and our results confirmed that they indeed stay at
this order of magnitude.
The extrapolated results are in good agreement with the
$Z$-factor predictions.

We conclude that the relevant perturbative contributions
to the  interface tension are not based on higher derivative terms
but are the effective potential and  the $\del\ph\del\ph$ term
as usually is assumed  in the literature.

The $Z$-factor in front
of the latter term has to be taken into account properly.
This can be done analytically using a semi-classical solution.
The values of the interface tension is lowered by this effect but
reliable calculations are not possible yet due to the strong gauge
fixing dependence of the one-loop $Z$-factor. A two-loop calculation of
this quantity may be helpful. Our results show
that including the corrections to the $\del\ph\del\ph$ term
is not sufficient to reach agreement with the published
lattice interface tension at intermediate Higgs masses.

In view of a lot of unexpected agreement between lattice data
and perturbative results and also some uncertainties in
particular in the case of the interface tension it would be desirable
to have a formalism bridging between perturbative treatment in the
Higgs phase and strong interaction dynamics in the hot phase.

\section{Acknowledgments}
We would like to thank Mikko Laine for useful discussions.

\end{document}